# Conductance Fluctuations and Domain Depinning in Quasi-2D Charge-Density-Wave 1$T$-TaS$_2$ Thin Films

Jonas O. Brown[1], Maedeh Taheri[1], Fariborz Kargar[1,2], Ruben Salgado[1], Tekwam Geremew[1], Sergey Rumyantsev[3], Roger K. Lake[1] and Alexander A. Balandin[1,2,4]*

[1]Department of Electrical and Computer Engineering, University of California, Riverside, California 92521 USA

[2]Department of Materials Science and Engineering, University of California, Los Angeles, California 90095 USA

[3]CENTERA Laboratories, Institute of High-Pressure Physics, Polish Academy of Sciences, Warsaw 01-142 Poland

[4]California NanoSystems Institute, University of California, Los Angeles, California 90095 USA

---

* Corresponding author (A.A.B.): balandin@seas.ucla.edu

## Abstract

We investigated the temperature dependence of the conductance fluctuations in thin films of the quasi-two-dimensional 1*T*-TaS$_2$ van der Waals material. The conductance fluctuations, determined from the derivative current-voltage characteristics of two-terminal 1*T*-TaS$_2$ devices, appear prominently at the electric fields that correspond to the transitions between various charge-density-wave macroscopic quantum condensate phases and at the onset of the depinning of the charge density wave domains. The depinning threshold field, $E_D$, *monotonically* increases with decreasing temperature within the nearly commensurate charge-density-wave phase. The $E_D$ value increases with the decreasing 1*T*-TaS$_2$ film thickness, revealing the surface pinning of the charge density waves. Our analysis suggests that depinning is absent in the commensurate phase. It is induced by the electric field but facilitated by local heating. The measured trends for $E_D$ of the domain depinning are important for understanding the physics of charge density waves in quasi-two-dimensional crystals and for developing electronic devices based on this type of quantum materials.

I. INTRODUCTION

Quasi-two-dimensional (2D) van der Waals materials are attracting growing interest for possible applications in electronics, optoelectronics, and energy conversion[1-3]. A particularly intriguing subgroup of quasi-2D materials is the one that reveals the charge-density-wave (CDW) quantum condensate phases[4-6]. The field of CDW materials and their properties experiences a rebirth owing to the emergence of new CDW materials of the quasi-2D transition-metal dichalcogenide (TMD) group and the extension of the quasi-one-dimensional (1D) CDW materials to the systems that reveal simultaneously CDW quantum condensate phases and topological effects[7-11]. Extra motivation comes from the discovery of quasi-2D and quasi-1D materials that reveal macroscopic CDW quantum phases above room temperature (RT)[11–13]. The 1T polymorph of TaS$_2$ (1$T$-TaS$_2$) is a prominent member of the TMD group, which undergoes several CDW quantum phase transitions[14-17]. Below ~200 K, 1$T$-TaS$_2$ is in the commensurate CDW (C-CDW) phase (see insets in Figure 1); above ~200 K, the C-CDW phase undergoes a transition to the nearly-commensurate CDW (NC-CDW) phase. This phase consists of the domains of the coherent C-CDW phase separated by the incommensurate CDW (IC-CDW) regions, which lack translation symmetry. At ~350 K, the NC-CDW phase of 1$T$-TaS$_2$ transitions to the IC-CDW phase; the material becomes a normal metal at ~550 K. The NC-CDW – IC-CDW phase transition is particularly interesting for practical applications of CDW quantum materials because it is above RT and near the temperature of the operation of conventional electronics. This phase transition can be induced or influenced by temperature, electrical bias, light, or other stimuli and it is usually accompanied by a hysteresis in the current-voltage (I-V) characteristics[18-24].

The functionality of the CDW devices can be based on different phenomena, including (i) the depinning and sliding of the CDW, with the resulting non-linear current increase and emergence of AC component in the output signal[25]; (ii) the Shapiro steps that appear under RF input signal synchronization with the sliding CDW[26]; (iii) hysteresis in *I-V* characteristics at the transition points between different CDW quantum phases, which can provide negative feedback for the oscillator function[11]. The CDW depinning and sliding have been utilized in "conventional" CDW devices based on materials with quasi-1D crystal structure[25,27,28]. The drawback of such devices was the requirement for a low temperature for their operation. The AC component under DC bias in quasi-1D materials, which revealed itself as the "narrow-band noise," was used to create RF oscillators[25]. The Shapiro steps in quasi-1D materials were shown to provide RF mixer functionality[26]. We have previously used the NC-CDW – IC-CDW phase transition in quasi-2D 1*T*-TaS$_2$ to demonstrate the voltage-controlled oscillator[11] and logic gates[29], operating at RT. One should keep in mind that the CDW depinning in quasi-2D CDW materials is fundamentally different from that in "conventional" quasi-1D CDW materials, which explains the scarcity of reports of its observation, let alone its device applications. This report is focused on elucidating the specifics of the CDW depinning in thin films of quasi-2D 1*T*-TaS$_2$.

In conventional theory for materials with quasi-1D crystal structure, defects pin the CDW to the lattice, requiring a finite threshold electric field, $E_T$, to de-pin the CDW and allow it to slide across in its IC-CDW phase. The result of the CDW depinning and sliding is an abrupt increase in current and AC signal[25]. The typical threshold fields for the depinning of CDWs in the quasi-1D materials vary from 40 mV/cm to 4 V/cm[30]. Since the C-CDWs are locked to the lattice, it was generally believed that sliding could only occur in the IC-CDW phase, and $E_T$ increases drastically at temperatures below the C-CDW – NC-CDW transition[25,27,28]. Most of the known materials revealed this trend. However, it was revealed that in K$_{0.3}$MoO$_3$, the threshold field for CDW depinning and sliding, $E_T$, decreases with the decreasing temperature even after the material was cooled below the C-CDW transition temperature[31]. It appeared that CDW depinning and sliding were possible even in the C-CDW phase. This discrepancy stimulated the search for new theoretical models for CDW transport in quasi-1D materials[32-35]. The situation with depinning in quasi-2D CDW materials is even more complicated. It was established that the depinning in 1*T*-TaS$_2$ does not result in a substantial current increase[12,36,37]. The process of CDW depinning in 1*T*-TaS$_2$ can be better conceptualized as the depinning of CDW domains, *i.e.*, the C-CDW domains in the NC-CDW phase become softer, looser, and, possibly, start rotating and changing in size at the depinning conditions[12,36,37]. This does not lead to the sliding of the entire CDW with the corresponding nonlinear increase in current. The domain depinning contributes to the conductance fluctuations, observed in the derivative *I-V*s and low-frequency noise[37], and allows for electrical gating of the CDW domains[36]. The reported depinning studies with electrical means in 1*T*-TaS$_2$ have been limited to RT[12,36,37].

To understand the physics of CDW transport in quasi-2D materials and assess the possibility of using depinning in 1*T*-TaS$_2$ for device functionality, it is important to study the threshold field for the depinning as a function of temperature. To emphasize that in quasi-2D 1T-TaS$_2$ we are dealing with the depinning of the CDW domains rather than with the depinning and sliding of the entire macroscopic CDW, we will use the symbol $E_D$ for the threshold field of the domain depinning. The questions we ask are: Does $E_D$ in quasi-2D 1T-TaS$_2$ increase with decreasing temperature like in NbSe$_3$ or decrease like in K$_{0.3}$MoO$_3$? Is there an abrupt change or continuous evolution in $E_D$ as the material enters the C-CDW phase? Can a CDW de-pin and, possibly, slide in the C-CDW phase? How does $E_D$ scale with the sample thickness? The answers to these questions can have a profound effect on understanding the physics of CDW phenomena in quasi-2D van der Waals materials and, at the same time, determine possible device applications of such quantum materials.

## II.     MATERIAL AND TEST STRUCTURE CHARACTERIZATION

In this study, we used high-quality commercial single-crystal 1T-TaS$_2$ as the source material. To verify the quality of the source material, particularly the presence of the NC-CDW – IC-CDW and C-CDW – NC-CDW phase transitions, we conducted differential scanning calorimetry (DSC) measurements. The DSC signal for the specific heat, $C_p$, with respect to temperature is presented in Figure 1 (a). One can clearly see the C-CDW – NC-CDW transition at ~216 K, the NC-CDW – IC-CDW transition at ~352 K, and the transition from the IC-CDW to the normal metal (NM) at ~531 K during the heating cycle. A small peak at 289 K is likely related to the intermediate T-phase inside the NC-CDW phase, which is often observed in the heating cycle[14,20]. To verify the CDW phase transitions even further, we measured the thermal diffusivity, $D$, and thermal conductivity, $\kappa$, of the 1T-TaS$_2$ as a function of temperature (Figure 1 (b)). The transition between the NC-CDW and IC-CDW states appears at ~350 K, close to the specific heat data, and the transition from IC-CDW to the NM phase emerges at ~540 K. Some variations in the transition temperatures can be attributed to the experimental techniques uncertainties and defects present in the "bulk" quantities of the materials.

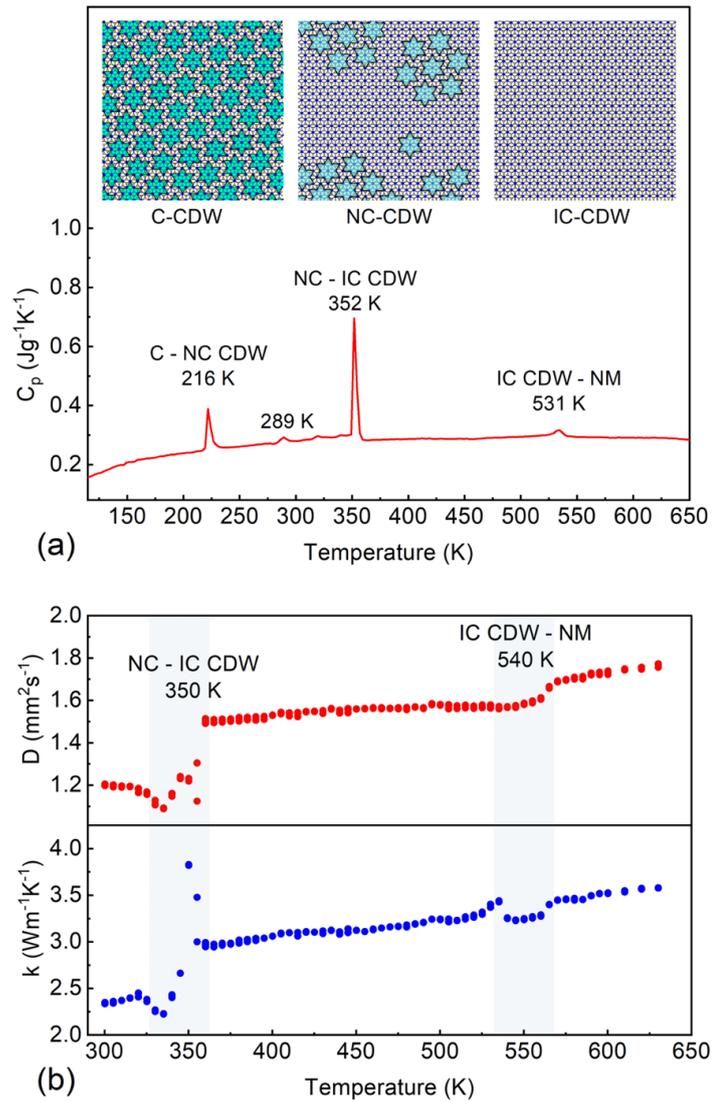

[**Figure 1:** Thermal characteristics of the as-synthesized 1*T*-TaS$_2$ material. (a) Specific heat showing the C-CDW to NC-CDW transition at 216 K and the NC-CDW to IC-CDW transition at 352 K and the IC-CDW to metallic transition at 531 K for a bulk sample of 1*T*-TaS$_2$. (b) Thermal diffusivity, *D*, and thermal conductivity, *κ*, of the as-synthesized 1*T*-TaS$_2$ material are presented in the top and bottom panels, respectively. The inset to panel (a) illustrates the three CDW phases in 1*T*-TaS$_2$. The commensurate CDW phase holds translational symmetry with the underlying lattice. The nearly commensurate CDW phase consists of commensurate CDW domains separated by a more metallic incommensurate CDW phase. The incommensurate CDW phase does not have translation symmetry; it has the highest electrical conductivity of the three phases.]

We prepared thin films of 1$T$-TaS$_2$ by mechanically exfoliating and placing them on a SiO$_2$/p$^+$-Si substrate using an in-house transfer system. To preserve the samples during the nanofabrication and avoid environmental exposure damage, a capping layer of thin $h$-BN was placed on top of the 1$T$-TaS$_2$ layers. The device structures were fabricated using electron-beam lithography (EBL). The atomic layer etching with SF$_6$ was used to expose 1$T$-TaS$_2$ film for the metal contacts. The contacts of Ti / Au (10 nm / 100 nm) were deposited using electron beam evaporation (EBE). An illustration of the crystal structure of 1$T$-TaS$_2$, a schematic of the test structures, *i.e.* devices, and optical microscopy images of representative devices are provided in the Supplementary Materials (see Figures S1 – S2).

The *I-V* characteristics of a representative device, with a channel thickness of $t \approx 30$ nm, are shown in Figure 2 (a). The transition from the NC-CDW phase to the IC-CDW phase is induced by applying a source-drain voltage to the 1$T$-TaS$_2$ channel. The electrical bias results in the current of normal, *i.e.*, individual, electrons. The shape and size of the hysteresis associated with the NC-CDW – IC-CDW transition are in line with prior reports[14,20,24,38]. The sweep rate in the measurements was 200 mV/s with the step of 2 mV. The direction of the voltage sweeps and the boundaries of the hysteresis at 100 K are indicated with the arrows and labeled as $V_H$ and $V_L$. The voltages $V_H$ and $V_L$ correspond to the electric bias that induces the NC-CDW to IC-CDW transition in the forward sweep and to the bias that induces the IC-CDW to NC-CDW transition in the reverse sweep, respectively. Looking at the data for other temperatures, we note that the hysteresis size shrinks as the temperature increases from 100 K to 300 K; the hysteresis disappears above $T = 350$ K, where the material enters the IC-CDW completely. Before the initial jump in the current in the hysteresis region, the *I-V* curve becomes super-linear due to Joule heating[11]. The local heating due to passing the current in the channel induces the NC-CDW – IC-CDW phase transition in these devices[14,20,24,38]. Naturally, the self-heating depends on the device design, size, substrate, and voltage sweeping rate. Inducing the CDW phase transitions by an electric field rather than local heating is also possible[14-16,36,39]. The fabricated 1$T$-TaS$_2$ devices were robust and the *I-V* characteristics were reproducible after several weeks of testing.

Examining Figure 2 (a) in more detail, we notice that the I-Vs for 100 K, 150 K and 200 K reveal only one NC-CDW – IC-CDW hysteresis and do not show features of C-CDW – NC-CDW hysteresis, which can be expected in the measurements conducted below the C-CDW –

NC-CDW transition temperature of ~225 K. There are two possible reasons for this. First, the 1*T*-TaS$_2$ films with a small channel thickness can be "locked" to the NC-CDW phase and do not undergo the transition to the C-CDW phase around ~225 K. This is a known phenomenon reported in several independent studies[14,20,40]. Second, in devices with a thickness in the range of ~ 4 nm – 50 nm fabricated on thermally resistive Si/SiO$_2$ substrates the self-heating effects can be significant. The local temperature of the device can be higher than that of the C-CDW – NC-CDW phase transition even if the I-Vs are measured at 100 K ambient temperature. In this case, one does not expect to see the C-CDW – NC-CDW transition. In the examined set of devices, we dealt with both situations. Figure 2 (b) shows the resistance as a function of temperature measured at low bias for devices with different channel thicknesses. One can see that depending on the thickness of the devices one can have a well-resolved hysteresis near the C-CDW – NC-CDW phase transition, a small one, or its absence. This is something that has to be considered in the data analysis.

Figures 2 (c) and (d) show the I-V and differential I-V characteristics, *dI/dV*, of another device at RT ($t \approx 30\ nm$). One can clearly identify the onset of conductance fluctuations at the bias voltage $V_D$=0.25 V and the large spikes in the conductance fluctuations, *i.e.* maximum of *dI/dV*, at the NC-CDW – IC-CDW phase transition. We use the subscript *D* in denoting the onset of the conductance fluctuations to emphasize that it is the CDW domain depinning. The large spike in Figure 2 (d) in the forward sweep at $V_H$=1.3 V exactly corresponds to the $V_H$ in Figure 2 (c). The transitions between various CDW quantum condensate phases are revealed both in I-V characteristics, as current or resistance jumps and hysteresis, and in the differential I-V characteristics, *dI/dV*, as the conductance fluctuations with the largest spikes. However, the domain depinning at $V_D$=0.25 is only observed in the differential characteristics in Figure 2 (d). No clear signatures are seen at this bias voltage in I-Vs, which are linear at this region (see Figure 2 (c)).

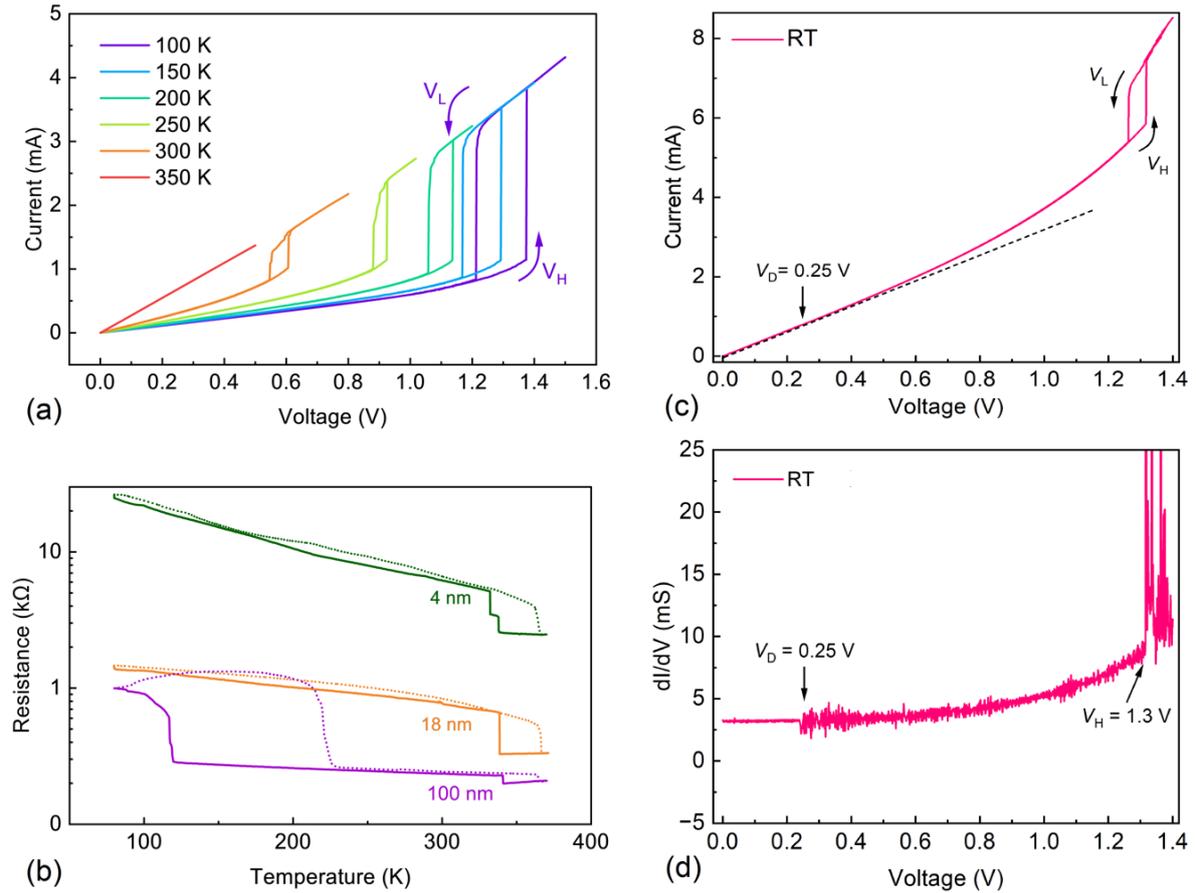

[**Figure 2**: Electrical characteristics of 1$T$-TaS$_2$ devices. (a) Current-voltage, I-V, characteristics of a representative device measured at different temperatures inside a cryostat. The characteristic hysteresis and resistivity change are induced by passing an electrical current *via* the device channel. The hysteresis is due to the transition between the NC-CDW and IC-CDW phases. (b) Resistance as a function of temperature measured at low bias for devices with different channel thicknesses. Note that the hysteresis associated with the NC-CDW – IC-CDW phase transition is present in all devices. The C-CDW – NC-CDW hysteresis appears clearly only in a thicker device. (c) Room-temperature I-V of a representative device, which shows the deviation from the linearity due to self-heating and hysteresis loop at the bias voltage of 1.2 V – 1.3 V due to the NC-CDW – IC-CDW phase transition. (d) Differential characteristics of the same device. Note the onset of conductance fluctuations, *dI/dV*, at $V_D$=0.25 V indicating the CDW domain de-pinning and large spikes in *dI/dV* at $V_H$=1.3 V corresponding to the NC-CDW – IC-CDW phase transition.]

### III. CHARGE DENSITY WAVE DEPINNING IN 2D MATERIALS

It was previously established that the CDW domain depinning at RT occurs at small biases where the *I-V* characteristics are linear[12,36,37]. The assessment of the local heating at the depinning points allowed us to conclude that the depinning is a field-induced phenomenon although it can be assisted by local heating[36,37]. As indicated above, the CDW domain depinning in 1$T$-TaS$_2$ should be understood as the commensurate CDW islands domains becoming softer and looser; possibly undergoing rotation, and altering their dimensions[12,36,37]. This process does not imply the CDW sliding as a whole, with a resulting large increase in current, owing to the "collective" current component, like is the case in quasi-1D CDW materials. For this reason, one cannot establish the threshold field for the CDW depinning and onset of sliding from the non-linear current increase. Rather, the CDW domain depinning in quasi-2D materials can be traced from differential I-V characteristics, *dI/dV*, as an on-set of conductance fluctuations, and the low-frequency noise measurements[12,37]. Let us now estimate the temperature rise at the CDW domain depinning point assuming that the local temperature is proportional to dissipated power, *i.e.* $\Delta T_H/\Delta T_D \sim V_H I_H/V_D I_D$, where $I_H$ and $I_D$ are the currents at $V_H$ and $V_D$, correspondingly. The temperature of the NC-CDW – IN-CDW phase transition is known as 350 K, giving the temperature rise at $V_H$ as $\Delta T_H$ ~50 K. Using the voltage and current values from Figure 2 (c), we obtained a rather small temperature rise of less than 4 K at the CDW domain depinning point, confirming the electric field nature of the depinning process in our devices.

We now proceed using the onset of the conductance fluctuations and large spikes in the conductance fluctuations (see Figure 2 (d)) to trace the CDW domain depinning and CDW phase transition fields at different temperatures. To investigate the dependence of $E_D$ on the temperature, we examine the differential I-V characteristics, focusing on the region, before the NC-CDW– IC-CDW phase transition. The testing protocol follows the sequence: At an initial cryostat temperature of 100 K, perform forward and reverse sweeps of the applied voltage and record the I-V and *dI/dV* responses; increase the temperature of the cryostat and repeat the prior steps until a final cryostat temperature is reached of approximately 350 K, where the hysteresis disappears. Figure 3 (a) – (f) shows representative snapshots of the conductance fluctuations in a 1$T$-TaS$_2$ device with a channel thickness of $t \approx 30$ nm as the function of electrical bias voltage at different measurement temperatures. Note that the actual temperature of the device channel can be different due to local heating. The temperatures indicated in the labels in Figure 3 (a) – (f) are the measurement temperatures, *i.e.* cryostat temperatures. Let us consider Figure 3 (a) panel with the data measured at 100 K in detail. One can clearly see the onset of the conductance fluctuations, *dI/dV*, at the electric bias of approximately 1.0 V. There is a large spike in *dI/dV* values at ~1.4 V. Comparing the resistive fluctuation data in Figure 3 (a) and I-V data for the case of 100 K in Figure 2 (a), one can see that the spike at ~1.4 V corresponds exactly to the $V_H$ bias voltage of the C-CDW – NC-CDW phase transition. The onset of the conductance fluctuations at ~1.0 V is associated with the CDW domain depinning. This is analogous to RT data presented in Figures 2 (c) and (d). The only difference is that when the measurements are done at low temperatures, the $V_D$ for the onset of fluctuations is substantially larger than that at RT. A much larger $V_D$ indicates that there is a significant temperature rise $\Delta T_D$ at this point. Using a simple estimate, as explained above, we obtain a temperature rise of above ~120 K, which suggests that even though the measurements are done at 100 K, the local temperature of the device channel is above ~220 K, and it is already in the NC-CDW phase. Thus, the onset of conductance fluctuations that we see at $V_D$=1.0 V corresponds to the domain depinning when the 1*T*-TaS$_2$ channel is in the NC-CDW phase.

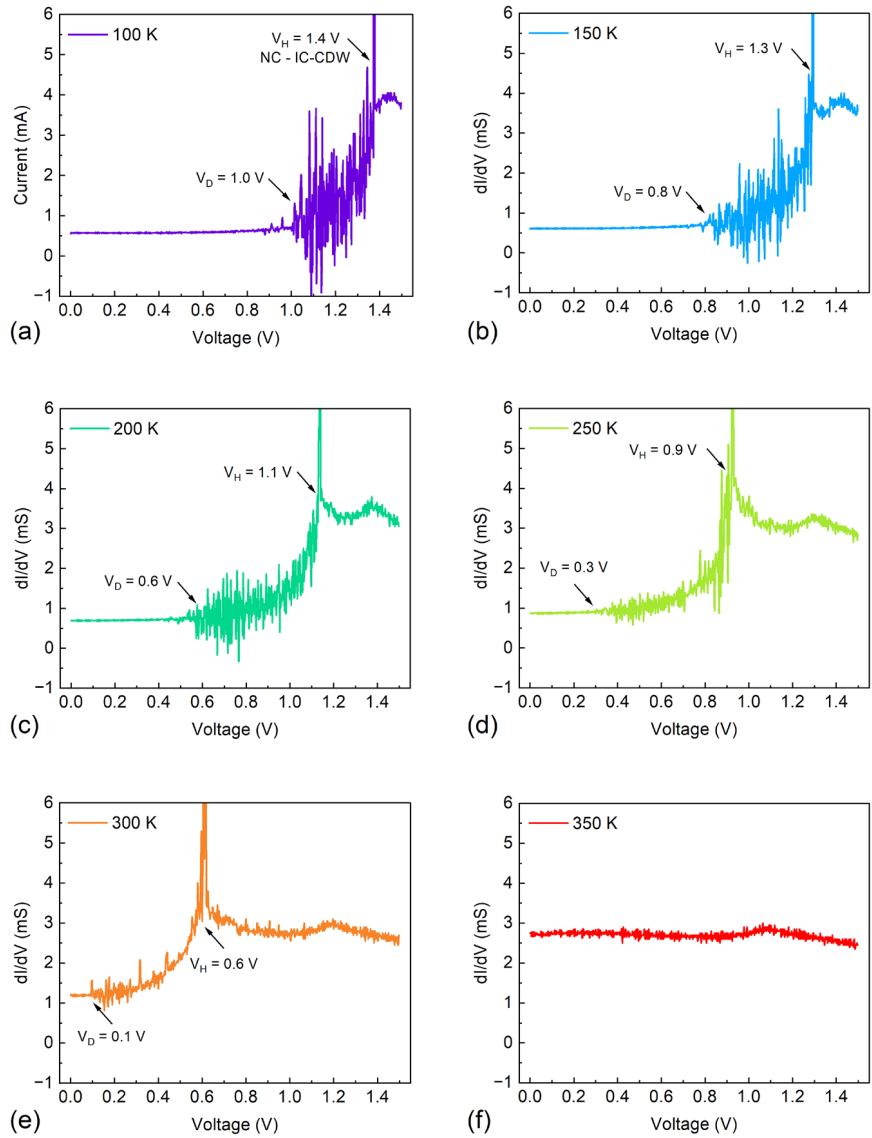

[**Figure 3:** Conductance fluctuations, *dI/dV*, as the function of the applied bias voltage for different measurement temperatures in the device with the channel thickness of ~30 nm. The data are shown for (a) 100 K, (b) 150 K, (c) 200 K, (d) 250 K, (e) 300 K, and (f) 350 K. The large spike at ~ 1.4 V for the device at 100 K corresponds to the NC-CDW – IC-CDW phase transition, labeled by $V_H$. The burst of the conductance fluctuations at smaller biases, $V_D$, indicates the CDW domain depinning. At higher temperatures, the onset of the conductance fluctuations and the voltage of the NC-CDW – IC-CDW phase transition shift to smaller bias voltages.]

The temperature evolution of the onset of the NC-CDW – IC-CDW phase transition, which can be extracted from the snapshots in Figure 3 (a – f), monitoring the large spike (indicated by the $V_H$), agrees with the trend one can trace in the I-Vs characteristics (see Figure 2 (a)) – the hysteresis shifts to smaller voltages and shrinks as the temperature increases. One can see that the onset of the conductance fluctuations moves to lower voltages as the measurement temperature increases, *e.g.* at 200 K, it starts approximately at ~0.6 V (Figure 3 (c)). It also becomes less pronounced, *e.g.* see Figure 3 (d) and (e). At 350 K, when the material is already in the IC-CDW phase the conductance fluctuations associated with the domain depinning and the phase transition disappear.

We now consider the conductance fluctuations in another representative device (*t*=100 nm) and start the measurements at T=80 K, instead of T=100 K. The snapshots for this device are presented in Figure 4 (a – f). Note that in this device, we first observe a large spike at a smaller bias voltage, followed by the burst of the conductance fluctuations, and then another spike at a larger bias voltage. The large spike at a smaller bias voltage corresponds to the C-CDW – NC-CDW phase transition while the spike at the larger bias corresponds to the NC-CDW – IC-CDW phase transition. One can see that the burst of conductance fluctuations, associated with the CDW domain depinning, always starts right after the C-CDW – NC-CDW phase transition. At T=80 K, this device with the thicker channel is in the C-CDW phase (see Figure 2 (b)). Passing the current we first induce the C-CDW – NC-CDW phase transition, after which, the CDW domain depinning occurs. This is different from the thinner channel device presented in Figure 3 (a-f), where the 1*T*-TaS$_2$ channel is in the NC-CDW phase and no traces of the large spike are observed before the onset of fluctuations associated with the CDW domain depinning. In Figure 4 (e), the local temperature of the channel is estimated to be above 214 K, which corresponds to the C-CDW – NC-CDW phase transition region. As a result, the spike is smaller in size and can be hardly recognized. At the measurement temperature T=180 K (Figure 4 (f)), the device is already in the NC-CDW phase, and we observe the domain depinning, without the initial spike, similar to that in Figure 3. Thus, the data presented in Figures 3 and 4 indicate that the domain depinning can only happen in the NC-CDW phase. It also suggests that the domain depinning itself is induced by the electric field but requires heating to occur when the material is initially in the C-CDW phase. Additional derivative I-V characteristics, supporting our conclusions, are provided in the Supplementary Materials (see Figures S3 – S5).

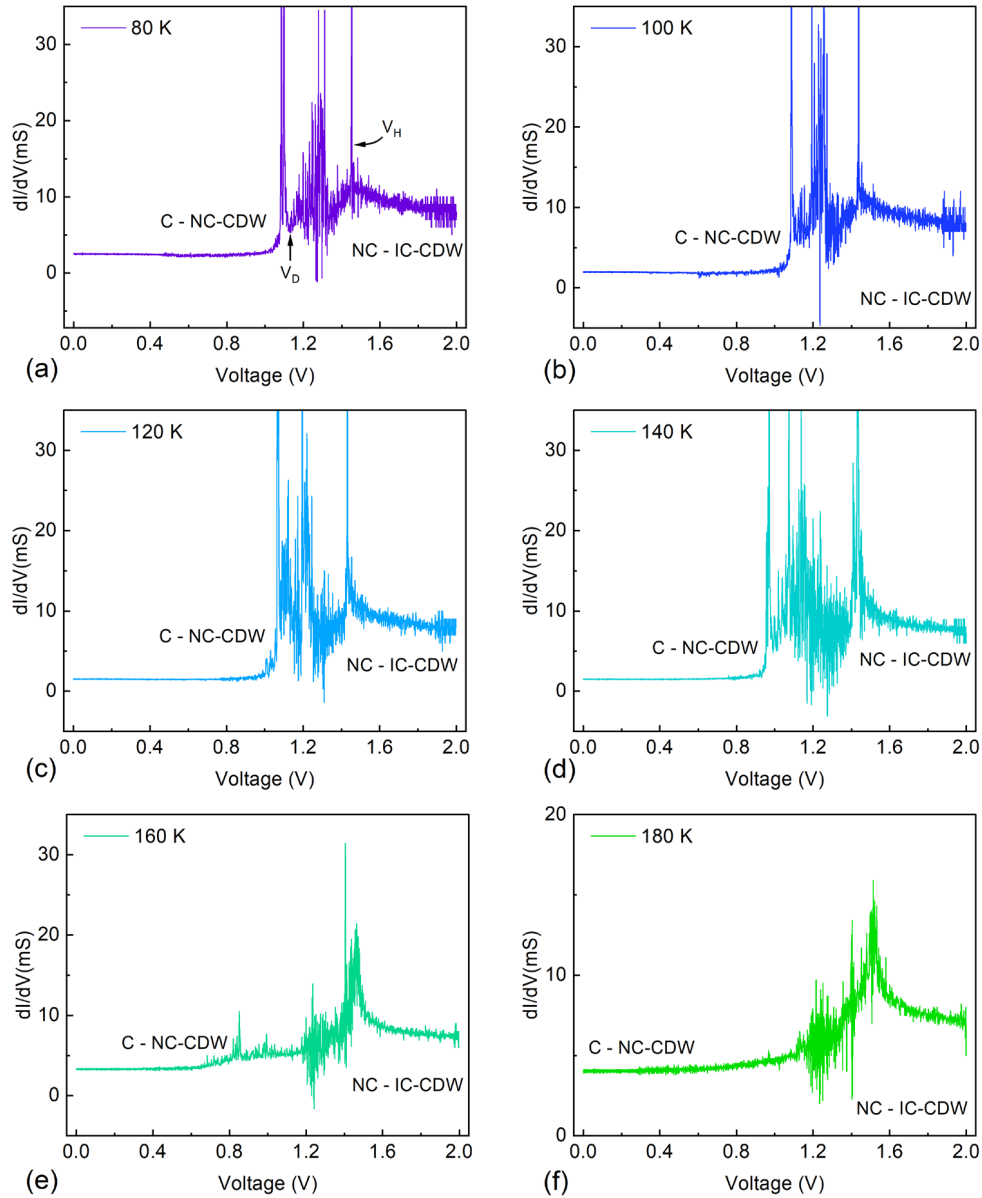

[**Figure 4:** Conductance fluctuations, *dI/dV*, as the function of the applied bias voltage for different measurement temperatures in the device with the channel thickness of ~100 nm. The data are shown for (a) 80 K, (b) 100 K, (c) 120 K, (d) 140 K, (e) 160 K, and (f) 180 K. In panels (a) - (d), one can see two large spikes and the burst of the conductance fluctuations between them. The large spike at a smaller bias voltage corresponds to the C-CDW – NC-CDW phase transition while the spike at the larger bias corresponds to the NC-CDW – IC-CDW phase transition. One can see that the burst of conductance fluctuations, associated with the CDW domain depinning, always starts right after the C-CDW – NC-CDW phase transition.]

To better visualize the general trend of the conductance fluctuations, we plotted the conductance fluctuations, *dI/dV*, as the function of both temperature and voltage using the contour map technique for the 30-nm thick device (see Figure 5). The maximum of the conductance fluctuations, *dI/dV*, which corresponds to the NC-CDW – IN-CDW phase transition can be traced as the deep red region with yellow highlights. The trend for the onset of the conductance fluctuations can be approximately traced along the black highlights, which are seen better than the changes in the shades of blue color. The uncertainty in the trend is relatively large due to the method, which relies on the derivatives I-V characteristics. Within the NC-CDW phase, the onset of conductance fluctuations corresponds to the domain depinning threshold, $V_D$. At lower temperatures, where the material is expected to be in the C-CDW phase, the onset of conductance fluctuations can indicate the C-CDW – NC-CDW phase transition accompanied by the domain depinning. We can estimate the effective threshold field for the CDW domain depinning, $E_D=V_D/L$, under the assumption of the linear voltage drop (*L* is the distance between the contacts for this device). For the measurement temperature *T*=250 K, where the material is inside the NC-CDW phase, the threshold field is in the range of ~ 1 kV/cm– 5 kV/cm, for the examined samples with different thicknesses, which is two to three orders of magnitude larger than the depinning threshold fields in the CDW materials with quasi-1D crystal structure[25,27,28,30,31,41,42]. This can be explained by the fact that the NC-CDW phase consists of C-CDW domains where the wave is locked within the domain borders. The depinning in quasi-2D material in the NC-CDW phase is more the "domain softening" rather than the depinning of the coherent CDW followed by its coherent sliding. The more metallic IC-CDW phase separating the C-CDW domains may also screen the electric field, making a linear voltage drop assumption to be too crude.

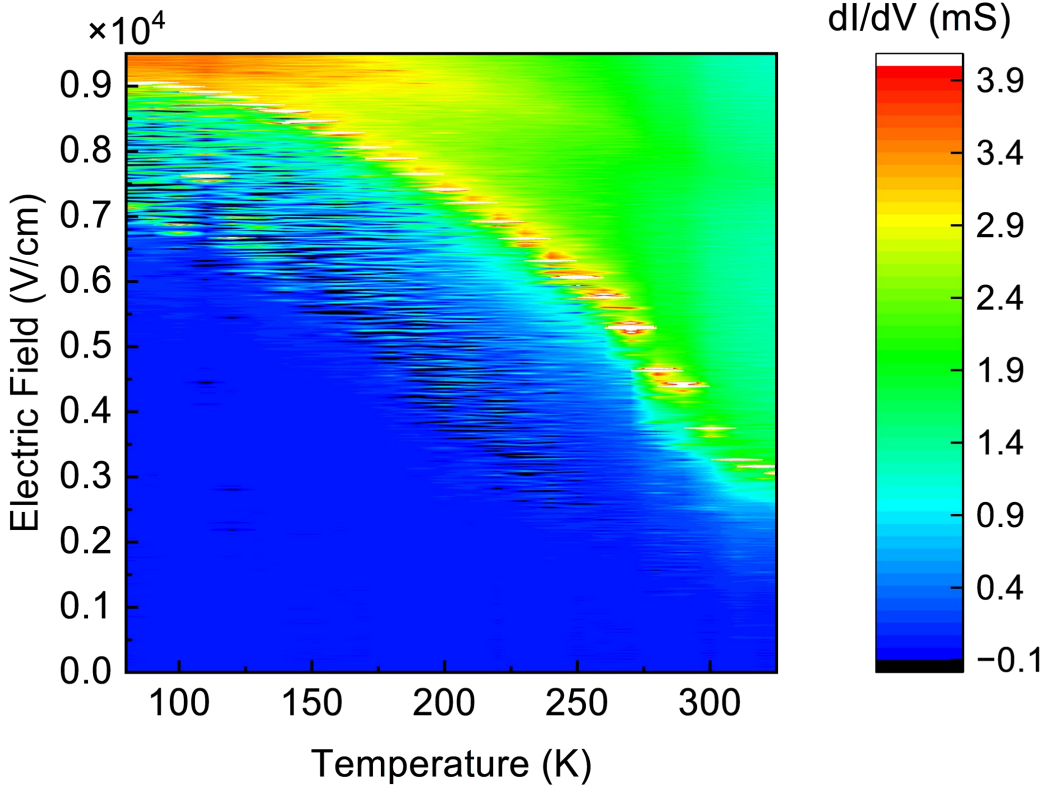

[**Figure 5:** Contour plot where the color represents the value of the conductance fluctuations, *dI/dV*. The scale is indicated next to the map. The data are shown in the cooling cycle for the device with a channel thickness of ~30 nm.]

### IV. TEMPERATURE AND SIZE EFFECTS ON CDW DEPINNING IN 2D MATERIALS

In Figure 6 (a), we present the domain depinning threshold field, $E_D=V_D/L$, as a function of the measurement temperature for several devices with different thicknesses. In all devices, we observe a nearly monotonic increase in $E_D$ with decreasing temperature. One should note that the devices with the thickness of the 1*T*-TaS$_2$ channel below approximately 45 nm were "locked" to the NC-CDW phase. The actual channel temperature was also higher than the measurement temperature indicated in Figure 6 (a) due to local heating. The device with the channel thickness $t$=100 nm at low temperature was in the C-CDW phase and had to be "unlocked" by self-heating to the NC-CDW phase to initiate the CDW depinning (see open symbols in Figure 6 (a) and 6 (b)). Using the above-described assumption that the local temperature is proportional to the dissipated power, *i.e. ΔT$_H$/ΔT$_D$~V$_H$I$_H$/V$_D$I$_D$*, we estimated the temperature rise due to the Joule heating and re-plotted the depinning threshold field as the function of the estimated channel temperature in Figure 6 (b). One can see that the trend for the monotonic increase in the depinning threshold with decreasing temperature is preserved. For comparison, in Figure 6 (c), we show the threshold field for the onset of the NC-CDW – IC-CDW phase transition with the corresponding well-known fitting with $(1- T/T_{\text{NC-IC}})^{1/2}$ trend, similar to the one observed in quasi-1D CDW materials[4].

The trend in the threshold field for the CDW domain depinning in quasi-2D material, presented in Figure 6 (a) and (b) is important information for further development of the theory of 2D CDW materials. It can be compared with the trends measured for CDW materials with quasi-1D crystal structures. In Figure 6 (d) we reproduced data points for NbSe$_3$[27] and K$_{0.3}$MoO$_3$[31]. The increasing threshold field with decreasing temperature below a certain temperature, $T_0$, which is near the CDW transition to the IC-CDW phase, was considered to be typical. Most of quasi-1D CDW materials reveal it, including NbSe$_3$, o-TaS$_3$, and organic materials[25,28]. Indeed, since the C-CDWs are locked to the lattice, it was generally believed that depinning and sliding could only occur in the IC-CDW phase. This point of view agreed well with the observed drastic increase in the threshold field at temperatures below the C-CDW transition, *e.g.*, in o-TaS$_3$[25]. However, in K$_{0.3}$MoO$_3$, the depinning threshold field continued decreasing after the materials were cooled below the C-CDW transition temperature (see Figure 6 (d)). This discrepancy stimulated the search for new theoretical models for CDW transport in quasi-1D materials. Among the proposed alternative approaches were the phase solitons[43], quantum fluidic solitons[44], and other approaches. These considerations explain the importance of the obtained results for a theoretical understanding of CDW transport in quasi-2D materials systems.

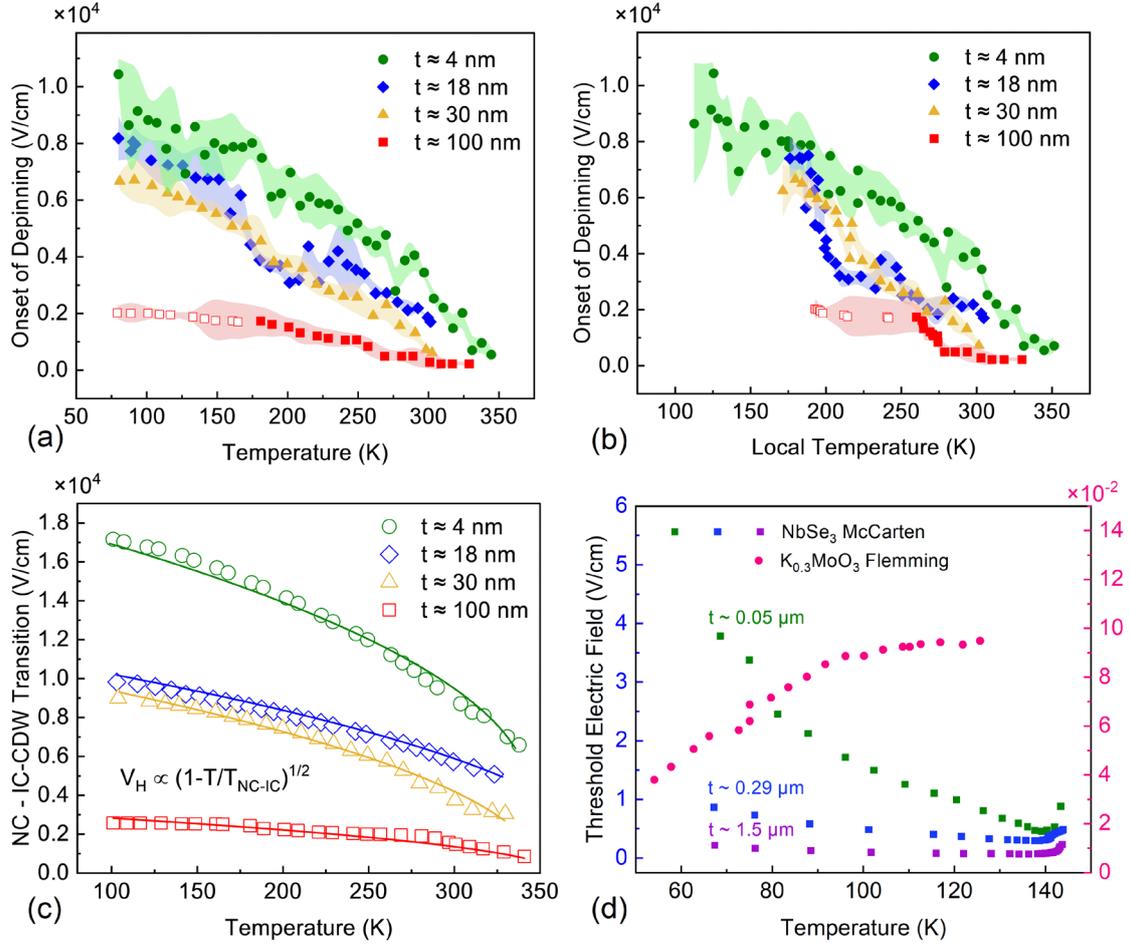

[**Figure 6:** Threshold field for the CDW domain depinning as a function of temperature. The data are presented for several devices with different thicknesses. The shaded regions represent the experimental uncertainty. (a) The CDW domain depinning field, $E_D$, as a function of the measurement temperature. (b) The CDW domain depinning field, $E_D$, as a function of the estimated local temperature of the $1T$-TaS$_2$ channel. (c) The field of the NC-CDW – IC-CDW phase transition as a function of the measurement temperature. (d) The threshold field of the CDW depinning in materials with the quasi-1D crystal structure. The data points are reproduced from Ref. [27] and [31]].

Another important observation from Figure 6 is that the values of the NC-CDW domain depinning threshold fields appear to have an inverse relation to the thickness of the films. This is consistent with the scaling behavior observed in quasi-1D IC-CDW materials[4,25,27]. In quasi-1D materials such as TaS$_3$ and NbSe$_3$, the threshold field, $E_T$, for IC-CDW sliding was found to scale as $\sim 1/\sqrt{A}$ where $A$ is the cross-sectional area of the material[6,45,46]. For a fixed width, this would result in the depinning field scaling as $\sim t^{-1/2}$, where $t$ is the thickness. Another study of the size dependence of the threshold voltage in NbSe$_3$, consistently found a $t^{-1}$ dependence for different samples with different doping concentrations and residual resistivities[27]. Furthermore, the $t^{-1}$ dependence was found to be independent of temperature. Other studies on the thickness dependence of the threshold field in o-TaS$_3$, found that it scaled as $t^{-1}$ for thicknesses $t > 60$ nm and $t^{-2/3}$ for the thicknesses $t < 60$ nm[47]. The conventional interpretation of the dependence of the depinning threshold field on the sample cross-sectional area or thickness is that it results from surface pinning of the CDW. The sample surface is "holding" the CDW, similar to the pinning by defects and extended defects. Since the nature of the NC-CDW domain depinning in quasi-2D 1$T$-TaS$_2$ is different from the IC-CDW depinning and sliding in materials with quasi-1D crystal structure one may expect somewhat different scaling with the thickness in our case. In Figure 7, we plot the threshold field for the CDW domain depinning, $E_D$, as a function of $1/\sqrt{A}$. This accounts for the variation of the width in some samples. The temperatures given in the legend are the estimated local temperatures due to the Joule heating. The plot shows that the depinning field in the NC-CDW phase also scales as $1/\sqrt{A}$. Further details of the fitting are provided in the SI. It should be noted that the older studies of depinning were performed over a thickness range of 0.1 μm – 10 $\mu$m[27], whereas the thicknesses that we consider are one to two orders of magnitude less.

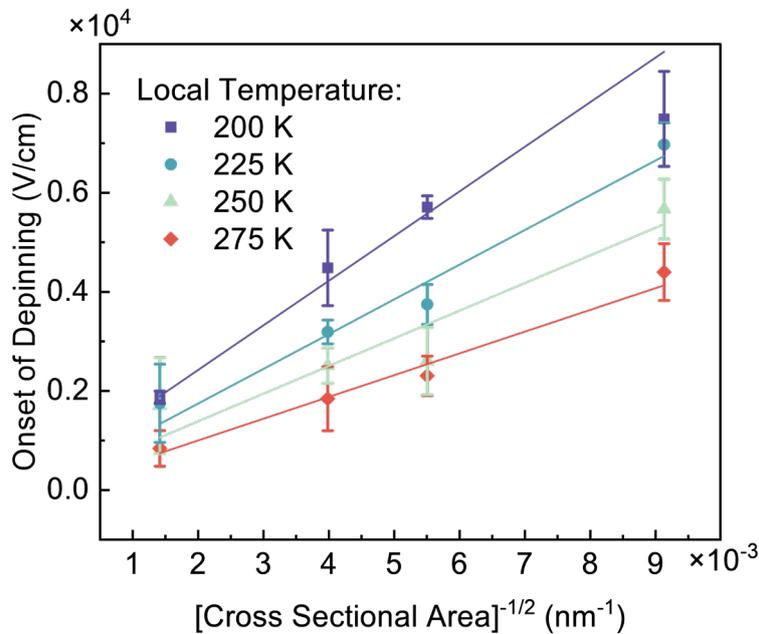

[**Figure 7:** Threshold field for the CDW domain depinning as a function of the cross-sectional area for different local device temperatures from 200 K to 275 K.]

V. CONCLUSIONS

We investigated the temperature dependence of the conductance fluctuations in thin films of the quasi-2D 1*T*-TaS$_2$ van der Waals material. The conductance fluctuations appear prominently at the electric fields, which correspond to the onset of the depinning of the charge density wave domains and the transitions between various CDW quantum condensate phases. The depinning threshold field monotonically increases with decreasing temperature as the material undergoes a transition from the nearly commensurate to the commensurate phase. The threshold field scales with the film thickness. The obtained results are important for understanding the physics of charge density waves in quasi-2D materials and for developing electronic devices based on this type of quantum materials.

**Acknowledgments**

The work at UC Riverside was supported, in part, by the U.S. Department of Energy Office of Basic Energy Sciences under contract No. DE-SC0021020 "Physical Mechanisms and Electric-Bias Control of Phase Transitions in Quasi-2D Charge-Density-Wave Quantum Materials". A.A.B. also acknowledges the Vannevar Bush Faculty Fellowship from the Office of Secretary of Defense (OSD), under the Office of Naval Research (ONR) contract N00014-21-1-2947. S.R. acknowledges partial support by the CENTERA Laboratories in the framework of the International Research Agendas program for the Foundation for Polish Sciences, co-financed by the European Union under the European Regional Development Fund (No. MAB/2018/9).

**Author Contributions**

A.A.B. conceived the idea, coordinated the project, contributed to experimental data analysis, and wrote the initial draft of the manuscript; J.O.B. fabricated devices in the cleanroom, measured *I-V* characteristics, and analyzed the experimental data; M.T. assisted with the device fabrication and testing; R.S. and T.G. conducted DSC measurements, and contributed to materials characterization. S.R., F.K., and R.K.L. contributed to the experimental data analysis. All authors contributed to the manuscript preparation.

**Supplemental Information**

The supplemental information is available at the journal website free of charge.

**Data Availability Statement**

The data that support the findings of this study are available from the corresponding author upon reasonable request.

# REFERENCES

[1] X. Li, L. Tao, Z. Chen, H. Fang, X. Li, X. Wang, J. B. Xu, and H. Zhu. Graphene and related two-dimensional materials: Structure-property relationships for electronics and optoelectronics. Applied Physics Reviews, 4(2):021306, 2017.

[2] C. C. Chiang, V. Ostwal, P. Wu, C.S. Pang, F. Zhang, Z. Chen, and J. Appenzeller. Memory applications from 2D materials. *Applied Physics Reviews*, 8(2):021306, 2021.

[3] C. S. Tang, X. Yin, and A. TS. Wee. 1D chain structure in 1T'-phase 2D transition metal dichalcogenides and their anisotropic electronic structures. *Applied Physics Reviews*, 8(1):011313, 2021.

[4] G. Grüner. Density waves in solids. *Addison-Wesley*, 1994.

[5] P. Monceau. Electronic crystals: an experimental overview. *Advances in Physics*, 61(4):325–581, 2012.

[6] S. V. Zaitsev-Zotov. Finite-size effects in quasi-one-dimensional conductors with a charge-density wave. *Physics-Uspekhi*, 47(6):533–554, 6 2004.

[7] A. A. Balandin, S. V. Zaitsev-Zotov, and G. Grüner. Charge-density-wave quantum materials and devices—New developments and future prospects. *Applied Physics Letters*, 119(17):170401, 2021.

[8] Y. A. Gerasimenko, I. Vaskivskyi, M. Litskevich, J. Ravnik, J. Vodeb, M. Diego, V. Kabanov, and D. Mihailovic. Quantum jamming transition to a correlated electron glass in 1T-TaS$_2$. *Nature Materials*, 18(10):1078--1083, 2019.

[9] I. Vaskivskyi, I. A. Mihailovic, S. Brazovskii, J. Gospodaric, T. Mertelj, D. Svetin, P. Sutar, and D. Mihailovic. Fast electronic resistance switching involving hidden charge density wave states. *Nature Communications*, 7(1):11442, 2016.

[10] M. J. Hollander, Y. Liu, W. Lu, L. Li, Y. Sun, J. A. Robinson, and S. Datta. Electrically Driven Reversible Insulator–Metal Phase Transition in 1T-TaS$_2$. *Nano Letters*, 15(3):1861–1866, 2015.

[11] G. Liu, B. Debnath, T. R. Pope, T. T. Salguero, R. K. Lake, and A. A. Balandin. A charge-density-wave oscillator based on an integrated tantalum disulfide-boron nitride-graphene device operating at room temperature. *Nature Nanotechnology*, 11(10):845–850, 2016.

[12] A. Mohammadzadeh, A. Rehman, F. Kargar, S. Rumyantsev, J. M. Smulko, W. Knap, R. K. Lake, and A. A. Balandin. Room temperature depinning of the charge-density waves in quasi-two-dimensional 1T-TaS$_2$ devices. *Applied Physics Letters*, 118(22):223101, 2021.

[13] A. Khitun, G. Liu, and A. A. Balandin. Two-dimensional oscillatory neural network based on room-temperature charge-density-wave devices. *IEEE Transactions on Nanotechnology*, 16(5):860–867, 2017.

[14] Y. Yu, F. Yang, X.F. Lu, Y. J. Yan, Y. H. Cho, L. Ma, X. Niu, S. Kim, Y. W. Son, D. Feng, S. Li, S. W. Cheong, X. H. Chen, and Y. Zhang. Gate-tunable phase transitions in thin flakes of 1T-TaS$_2$. *Nature Nanotechnology*, 10(3):270–276, 2015.

[15] J. Ravnik, M. Diego, Y. Gerasimenko, Y. Vaskivskyi, I. Vaskivskyi, T. Mertelj, J. Vodeb, and D. Mihailovic. A time-domain phase diagram of metastable states in a charge ordered quantum material. *Nature Communications*, 12(1):2323, 2021.

[16] A. Mraz, R. Venturini, D. Svetin, V. Sever, I.A. Mihailovic, I. Vaskivskyi, B. Ambrozic, M., G. Dražić, D. Stornaiuolo, F. Tafuri, D. Kazazis, J. Ravnik, Y. Ekinci, and D. Mihailovic. Charge Configuration Memory Devices: Energy Efficiency and Switching Speed. 41:3, 2022.

[17] J. Vodeb, V. V. Kabanov, Y. A. Gerasimenko, R. Venturini, J. Ravnik, M. A. van Midden, E. Zupanic, P. Sutar, and D. Mihailovic. Configurational electronic states in layered transition metal dichalcogenides. *New Journal of Physics*, 21(8):083001, 2019.

[18] J. A. Wilson, F. J. Di Salvo, and S. Mahajan. Charge-density waves and superlattices in the metallic layered transition metal dichalcogenides. *Advances in Physics*, 24(2):117–201, 1975.

[19] A. W. Tsen, R. Hovden, D. Wang, Y. D. Kim, K. A. Spoth, Y. Liu, W. Lu, Y. Sun, J. C. Hone, L. F. Kourkoutis, P. Kim, and A. N. Pasupathy. Structure and control of charge density waves in two-dimensional 1T-TaS$_2$. *Proceedings of the National Academy of Sciences of the United States of America*, 112(49):15054–15059, 2015.

[20] B. Sipos, A. F. Kusmartseva, A. Akrap, H. Berger, L. Forró, and E. Tutî. From Mott state to superconductivity in 1T-TaS$_2$. *Nature materials*, 7(12):960–965, 2008.

[21] A. K. Geremew, S. Rumyantsev, F. Kargar, B. Debnath, A. Nosek, M. A. Bloodgood, M. Bockrath, T. T. Salguero, R. K. Lake, and A. A. Balandin. Bias-Voltage Driven Switching of the Charge-Density-Wave and Normal Metallic Phases in 1T-TaS$_2$ Thin-Film Devices. *ACS Nano*, 13(6):7231–7240, 2019.

[22] T. Patel, J. Okamoto, T. Dekker, B. Yang, J. Gao, X. Luo, W. Lu, Y. Sun, and A.W. Tsen. Photocurrent Imaging of Multi-Memristive Charge Density Wave Switching in Two-Dimensional 1T-TaS$_2$. *Nano Letters*, 20(10):7200–7206, 2020.

[23] G. Liu, E. X. Zhang, C. D. Liang, M. A. Bloodgood, T. T. Salguero, D. M. Fleetwood, and A. A. Balandin. Total-Ionizing-Dose Effects on Threshold Switching in 1T-TaS$_2$ Charge Density Wave Devices. *IEEE Electron Device Letters*, 38(12):1724–1727, 2017.

[24] A. K. Geremew, F. Kargar, E. X. Zhang, S. E. Zhao, E. Aytan, M. A. Bloodgood, T. T. Salguero, S. Rumyantsev, A. Fedoseyev, D. M. Fleetwood, and A. A. Balandin. Proton-irradiation-

immune electronics implemented with two-dimensional charge-density-wave devices. *Nanoscale*, 11(17):8380–8386, 2019.

[25] G. Grüner, A. Zettl, W. G. Clark, and A. H. Thompson. Observation of narrow-band charge-density-wave noise in TaS$_3$. *Physical Review B*, 23(12):6813–6815, 1981.

[26] A. Zettl and G. Grüner. Observation of shapiro steps in the charge-density-wave state of NbSe$_3$. *Solid state communications*, 46(7):501–504, 1983.

[27] J. McCarten, D. A. DiCarlo, M. P. Maher, T. L. Adelman, and R. E. Thorne. Charge-density-wave pinning and finite-size effects in NbSe$_3$. *Physical Review B*, 46(8):4456–4482, 1992.

[28] L. Forro, R. Lacoe, S. Bouffard, and D. Jérome. Defect-concentration dependence of the charge-density-wave transport in tetrathiafulvalene tetracyanoquinodimethane. *Physical Review B*, 35(11):5884–5886, 1987.

[29] A. G. Khitun, A. K. Geremew, and A. A. Balandin. Transistor-Less Logic Circuits Implemented with 2-D Charge Density Wave Devices. *IEEE Electron Device Letters*, 39(9):1449–1452, 2018.

[30] Z. Z. Wang, P. Monceau, H. Salva, C. Roucau, L. Guemas, and A. Meerschaut. Charge-density-wave transport above room temperature in a polytype of NbS$_3$. *Physical Review B*, 40(17):11589, 1989.

[31] R. M. Fleming, L. F. Schneemeyer, and D. E. Moncton. Commensurate incommensurate transition in the charge-density-wave state of K$_{0.30}$MoO$_3$. *Physical Review B*, 31(2):899–903, 1985.

[32] J. Bardeen. Theory of non-Ohmic conduction from charge-density waves in NbSe$_3$. *Physical Review Letters*, 42(22):1498–1500, 1979.

[33] J. Bardeen. Basis for tunneling theory of charge-density wave depinning. *Zeitschrift für Physik B Condensed Matter*, 67(4):427–433, 1987.

[34] K. Yackoboski, Y. H. Yeo, G. C. McGonigal, and D. J. Thomson. Molecular position at the liquid/solid interface measured by voltage-dependent imaging with the STM. *Ultramicroscopy*, 42:963–967, 1992.

[35] L. Pietronero and M. Versteeg. Theory of the threshold field for the depinning transition of a charge density wave. *Physica A: Statistical Mechanics and its Applications*, 179(1):1–15, 1991.

[36] M. Taheri, J. Brown, A. Rehman, N. Sesing, F. Kargar, T. T. Salguero, S. Rumyantsev, and A. A. Balandin. Electrical Gating of the Charge-Density-Wave Phases in Two-Dimensional h-BN/1T-TaS$_2$ Devices. *ACS Nano*, 16(11):18968–18977, 2022.

[37] G. Liu, S. Rumyantsev, M. A. Bloodgood, T. T. Salguero, and A. A. Balandin. Low-Frequency Current Fluctuations and Sliding of the Charge Density Waves in Two-Dimensional Materials. *Nano Letters*, 18(6):3630–3636, 2018.

[38] L. Stojchevska, I. Vaskivskyi, T. Mertelj, P. Kusar, D. Svetin, S. Brazovskii, and D. Mihailovic. Ultrafast switching to a stable hidden quantum state in an electronic crystal. *Science (New York, N.Y.)*, 344(6180):177–180, 2014.

[39] C. Zhu, Y. Chen, F. Liu, S. Zheng, X. Li, A. Chaturvedi, J. Zhou, Q. Fu, Y. He, Q. Zeng, H.J. Fan, H. Zhang, W.J. Liu, T. Yu, and Z. Liu. Light-Tunable 1T-TaS$_2$ Charge-Density-Wave Oscillators. *ACS Nano*, 12(11):11203–11210, 2018.

[40] Y. D. Wang, W. L. Yao, Z. M. Xin, T. T. Han, Z. G. Wang, L. Chen, C. Cai, Yuan Li, and Y. Zhang. Band insulator to Mott insulator transition in 1T-TaS$_2$. *Nature Communications*, 11(1):1–7, 2020.

[41] R. M. Fleming. Electric-field depinning of charge-density waves in NbSe$_3$. *Physical Review B*, 22(12):5606–5612, 1980.

[42] E. B. Lopes, M. J. Matos, R. T. Henriques, M. Almeida, and J. Dumas. CDW dynamics in the quasi-one-dimensional molecular conductors (Per)$_2$M(mnt)$_2$, (M=Au and Pt). *Synthetic Metals*, 86(1-3):2163–2164, 1997.

[43] B. Horovitz and J. A. Krumhansl. Solitons in the peierls condensate: Phase solitons. *Physical Review B*, 29(4):2109, 1984.

[44] J. H. Miller and M. Y. Suárez-Villagrán. Quantum fluidic charge density wave transport. *Applied Physics Letters*, 118(18):184002, 2021.

[45] D. V. Borodin, S.V. Zaitsev-Zotov, and F.Y. Nad. Coherence of a charge density wave and phase slip in small samples of a quasi-one-dimensional conductor TaS$_3$. *Zh. Eksp. Teor. Fiz*, 93:1394–1409, 1987.

[46] G. Grüner. The dynamics of charge-density waves. *Reviews of Modern Physics*, 60(4):1129, 1988.

[47] E. Slot. Microscopic Charge Density Wave Transport. *Leiden University*, Ph.D. (Diss):111, 2005.